\documentclass[aip,rsi,amsmath,amssymb,preprint]{revtex4-1}

\usepackage{graphicx,textcomp}
\usepackage{dcolumn,mathrsfs}
\usepackage{tabularx}
\usepackage{bm}

\begin{document}

\preprint{AIP/123-QED}

\title[Universality class change due to angle of deposition]{Universality class change due to angle of deposition of thin-film growth by random particles aggregation}

\author{A. C. A. Vilas Boas}
\author{F. L. Forgerini}
 \email{fabricio.forgerini@ufsb.edu.br}
 \affiliation{Campus Paulo Freire, Universidade Federal do Sul da Bahia\\
45988-058 \hspace{5mm} Teixeira de Freitas - BA \hspace{5mm} Brazil}

\date{\today}

\begin{abstract}
In this work we study numerically the effects of the angle of deposition of particles in the growth process of a thin-film generated by aggregation of particles added at random. The particles are aggregated in a random position of an initially flat surface and with a given angle distribution. This process gives rise to a rough interface after some time of deposition. We performed Monte Carlo simulations and, by changing the angle of deposition, we observed a transition from the random deposition (RD) universality class to the Kardar-Parisi-Zhang (KPZ) universality class. We measured the usual scaling exponents, namely, the roughness ($\alpha$), the growth ($\beta$) and the dynamic ($z$) exponents. Our results show that the particles added non-perpendicularly to the substrate, can change the universality class in a discrete atomistic random deposition model. When particles are deposited with an angle of $45^{\circ}$ in relation to the surface, the same values of the Ballistic Deposition model are observed in the Random Deposition model. We also propose an analytic approach, using a differential stochastic equation to analyze the growth process evolution, and our theoretical results corroborate the computer simulations.
\end{abstract}


\maketitle

\section{Introduction}
\label{Introduction}

The understanding of physical process that take place at surfaces and interfaces has attracted interest of researchers from different fields \cite{barabasi, meakin}. Specially motivated by the technological applications developed from thin-films \cite{huang, Forgerini_Marchiori}, the investigation on the formation of structures due to the deposition of atoms or particles has been the subject of large number of recent studies both experimental and theoretical \cite{Ebrahiminejad, Kolakowska, fabio_reis, Mirabella}.

By controlling the surface processes, one can control physical, chemical, optical and mechanical properties of the material, leading to the development of new devices with practical purposes \cite{livro:barkema}. Theoretical and computational models represent a powerful tool to study the growth of thin-films and interfaces, where the physicists can apply the well-known methods of statistical physics to describe these non-equilibrium phenomena.

New advances in recent years allowed a better understanding of the fundamental phenomena which govern the deposition of particles forming a thin film at nanoscale. Atomistic models have been largely applied in this field of study, using different forms of particles \cite{Ebrahiminejad, forgerini_figueiredo1, forgerini_figueiredo2}. Although many models are quite simpler, one can use them as a good starting point to study more sophisticated processes that are directly related to the experimental growth process and techniques.

New experimental techniques, such as sputtering or Molecular Beam Epitaxy (MBE), can provide suitable materials for a large range of applications, such as medicine \cite{Karagkiozaki}, in which thin and ultra-thin film coatings for stent devices are, perhaps, one of most remarkable examples of nanostructured biomaterials. In another field, the electronic and nano electronic industry have an increasing demand for devices at nanoscale, in which the surface morphology play a very important role for applications in solar cells \cite{Karmhag}, information storages \cite{Nielsch} and carbon nanotubes \cite{Che}.

In this paper we introduce a modification in the Random Deposition (RD) model, considering that the particles can be aggregated with different angles in relation to the initially flat substrate. Due to the simplicity of the model, random deposition is widely studied, but the uncorrelated interface resulting from this process is not realistic. In order to make the model more realistic and applicable to the surfaces science, we modify the RD model and introduce a natural correlation on the surface by adding particles obliquely. We study our model by means of computer simulations and stochastic growth equations and we show that, by adding particles obliquely to the surface, the RD model can produce the same scaling exponents of the Ballistic Deposition (BD) model when the angle of deposition is 45\textdegree~ in relation to the initially flat substrate.

This paper is organized as follows: In Sec~\ref{Methodology} we analyze the standard methods for theoretical analysis of thin-films. In Sec.~\ref{Model} we present our model, the deposition rules and simulations details. The discussion of the results of the numerical simulations and the theoretical analysis of the stochastic equations is presented in Sec.~\ref{Results}, and the main conclusions are presented in Sec.~\ref{Conclusions}.

\section{Methodology for computer simulation and theoretical analysis}
\label{Methodology}

In the field of theoretical surface growth there are a few standard tools developed for the analysis of surfaces and interfaces. One method of analysis of surface growth is through scaling concepts. There are some characteristics of surfaces and interfaces which obey some scaling relations. Studying these relations and their corresponding exponents, one can define a few universality classes in which different processes share the same scaling behavior \cite{barabasi, meakin}.

Another possible way to study theoretically these processes is through continuum growth equations. Stochastic differential equations describe the interface at large length scales. One can associate a specific growth process with an equation which classify them into the proper universality class. The Random Deposition process is the simplest discrete atomistic model and can be described by the equation

\begin{equation}
 \frac{\partial h(\vec{r}, t)}{\partial t} = F + \eta(\vec{r}, t),
 \label{stochastic_eq}
\end{equation}
where F represents the average number of particles per unit time that are added to the substrate at given position and $\eta(\vec{r}, t)$ represents the random fluctuation of this process, a noise that does not show spatial correlation in the substrate.

On the other hand, one of the most important continuum growth equations, related to the BD model of particles, represents a wide variety of processes of surface growth and non-equilibrium interfaces, such as those related to the formation of porous surfaces, corrosion processes of metallic surfaces and dissolution of a crystalline solid in a liquid medium \cite{freis, fernando-oliveira}. The equation \ref{KPZ}, called KPZ, has the form,

\begin{equation}
 \frac{\partial h(\vec{r}, t)}{\partial t} = \nu \triangledown^2 h + \frac{\lambda}{2} (\triangledown h)^2 + \eta(\vec{r}, t),
 \label{KPZ}
\end{equation}
and includes a nonlinear term $\frac{\lambda}{2} (\triangledown h)^2$, that takes into account lateral growth, describing the aggregation of particles parallel to the surface. Unfortunately, in many cases, continuum growth equations do not have and exact solutions for a specific class of problems or a specific spatial dimension, restricting their use.

For the class of problems previously mentioned, one can analyze them by means of discrete growth models, where the deposition process is simulated by a computer algorithm and the surface morphology is reproduced. Simulations are an essential link between theory and experiments and can provide some morphological details that are usually neglected by the equations but revealed by experimental techniques \cite{livro:landau}.

In order to study numerically the morphology of a surface, one can calculate the interface width (surface roughness), $w(L,t)$, a function of time and the linear size of the substrate. In order to calculate $w(L,t)$, we determine the vertical height of the surface relative to the substrate at a given time $t$, $h(\vec{r}, t)$, where $\vec{r}$ gives the position on the substrate. The roughness $w(L,t)$ is defined as the mean square fluctuation of the height, $w(L,t) = \langle [h(\vec{r}, t) - \bar{h}(t)]^2\rangle^{1/2}$ where $\bar{h}(t)$ is the average value of the surface’s height at a given instant of time $t$.

\begin{figure}[!htb]
\includegraphics[width=0.8\linewidth]{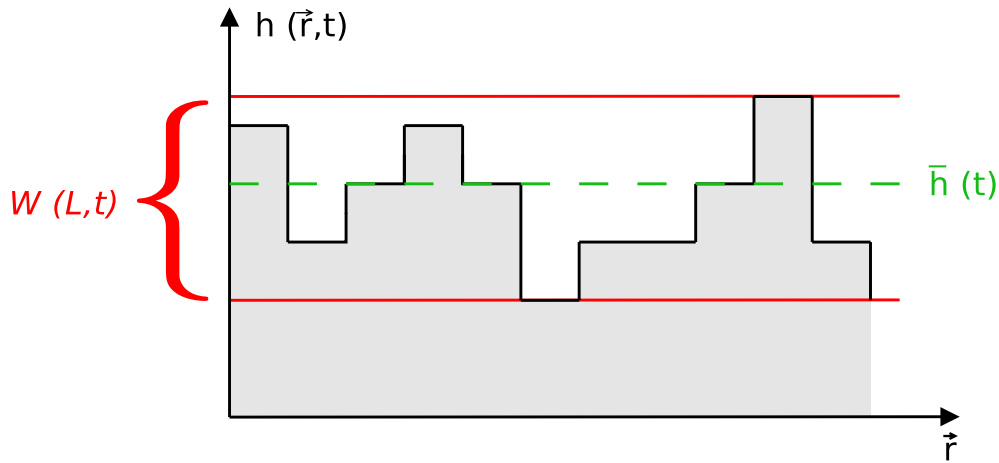}
\caption{Graphic representation of the cross section of a surface. One can see in the figure the average height $\bar{h}(t)$ and the definition of surface roughness $w(L,t)$.}
\label{representation}
\end{figure}

Figure~\ref{representation} represents a schematic deposition, in (1+1) dimensions, in which (1+1) means one spatial dimension plus one-time dimension. One can see that $\bar{h}(t)$ and $w(L,t)$ are also represented.

In both cases, by scaling concepts, one can study and characterize a growth model that represents in some sense a real surface growth process. The surface roughness increases as a power of time initially, $w(L,t) \sim t^\beta$, and, after some time of deposition, $t_x$, the roughness saturates, $w_{sat}(L) \sim L^\alpha$. The time necessary to saturation depends on the system size, $t_x \sim L^z$. These exponents are not independent and they are related in the form $z = \alpha/\beta$ \cite{family_vicsek}. In theoretical studies of surface growth, one is interested in the calculation of these scaling exponents.

\section{A Model for angular particle aggregation}
\label{Model}

In the present study, particles with size of one lattice unit are randomly dropped over the initially flat substrate. The difference between our model and the classic Random Deposition Model is that in our model particles are added obliquely to the substrate, in different angles of deposition, introducing a natural correlation among the first neighbors, reflecting a lateral growth of the surface.

The particles are released from a randomly chosen position, far from the surface, with a desired angle in relation to the substrate, following a trajectory until reaches the surface, whereupon they are deposited in the landing point. In order to make the model closer to real deposition processes, such as the colloidal aggregation or the vapor deposition, we introduce a deviation of $\pm 10\%$ in the values of the deposition angle, which defines the trajectories of the particles. 

\begin{figure}[!htb]
\includegraphics[width=0.6\linewidth]{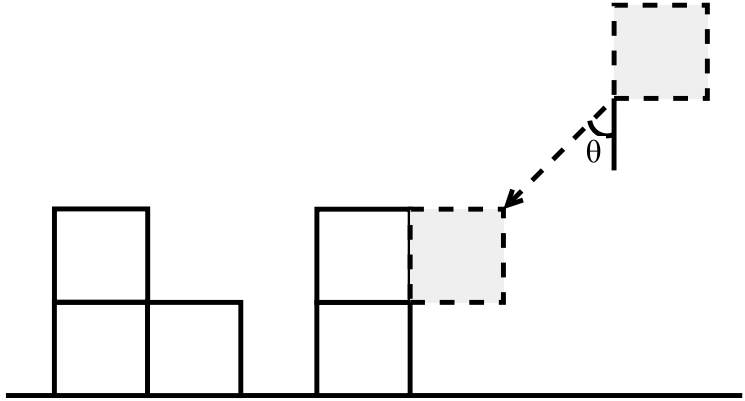}
\caption{Illustration of the growth process of our model where particles are dropped with an angle $\theta$.}
\label{fig_model}
\end{figure}

A schematic representation of our model in figure~\ref{fig_model}, in which one can see the aggregation of particles leading to a lateral surface growth. This lateral growth is observed in real growth processes such as the electrochemical deposition and it is also observed in corrosion and oxidation processes in which the correlations among positions on the surface play an important role on the complex surface morphology.

We perform Monte Carlo simulations to study the surface growth in a variety of linear lattices sizes and angles of deposition. For the computer simulations, we define one Monte Carlo step, the unit of time in this problem, as the time required to deposit L particles on the substrate. In this study, we define that particles are not allowed to diffuse after the deposition.

\section{Results and discussion}
\label{Results}

\begin{figure}[!htb]
\reflectbox{\includegraphics[width=0.4\linewidth]{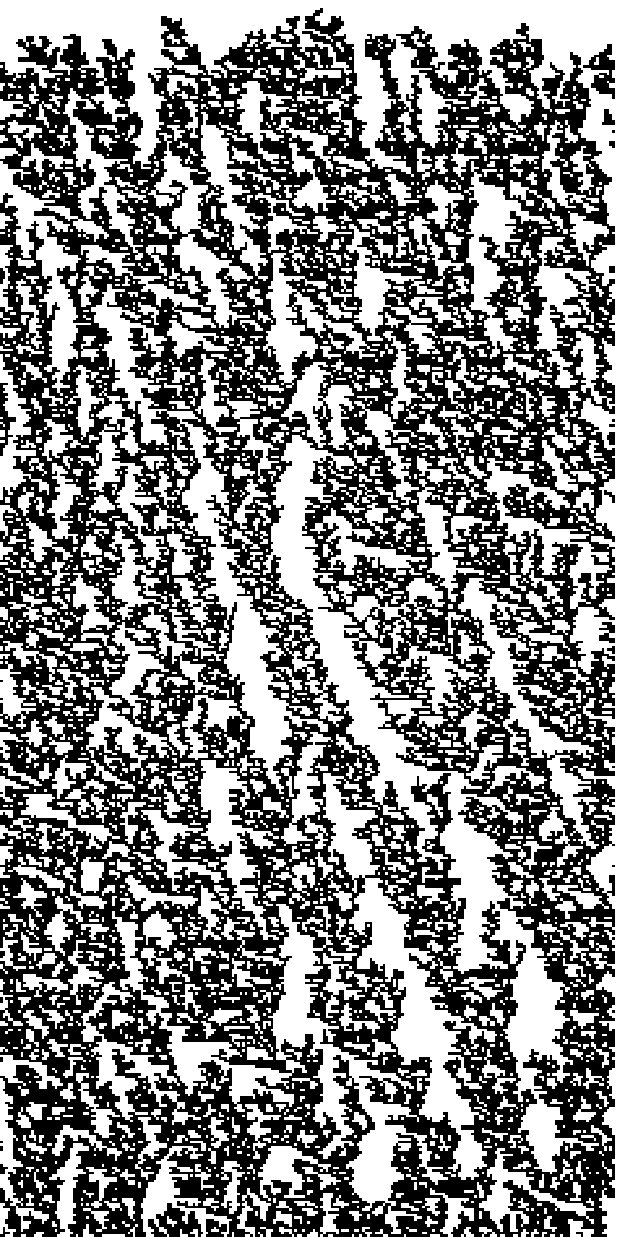}}
\caption{Graphic representation of the deposition process in one dimension and linear size $L = 1024$. The particles are dropped with angle $\theta = 45$\textdegree.}
\label{deposition}
\end{figure}

Our Monte Carlo simulations were performed on squared lattices with linear size ranging from L~=~128 to 4096 and with different angles of deposition between 0~\textdegree~$\leq~\theta~\leq$~45\textdegree. At initial time steps, the growth is close to a surface generated by a RD model. However, as the time goes by, the lateral growth take place and the surface morphology change drastically. A cross section of a surface generated by our simulation in shown in the figure \ref{deposition}, as one can see that the lateral aggregation of particles as the growth process evolve.

From our simulations, we obtained the value of the exponents $\alpha$ and $\beta$ for different values of $\theta$. The roughness exponent $\alpha$, in our model, do not depends on the angle of deposition for $\theta \geq 20$\textdegree. The best value of $\alpha$ can be estimated from the $W_{sat}$ after the extrapolation the effective exponents defined by the equation\cite{fabio_euzebio}
\begin{equation}
\alpha(L) \equiv \frac{ln[w_{sat}(L) / w_{sat}(L/2)]}{ln 2}.
\end{equation}

\begin{figure}[!htb]
\includegraphics[width=0.8\linewidth]{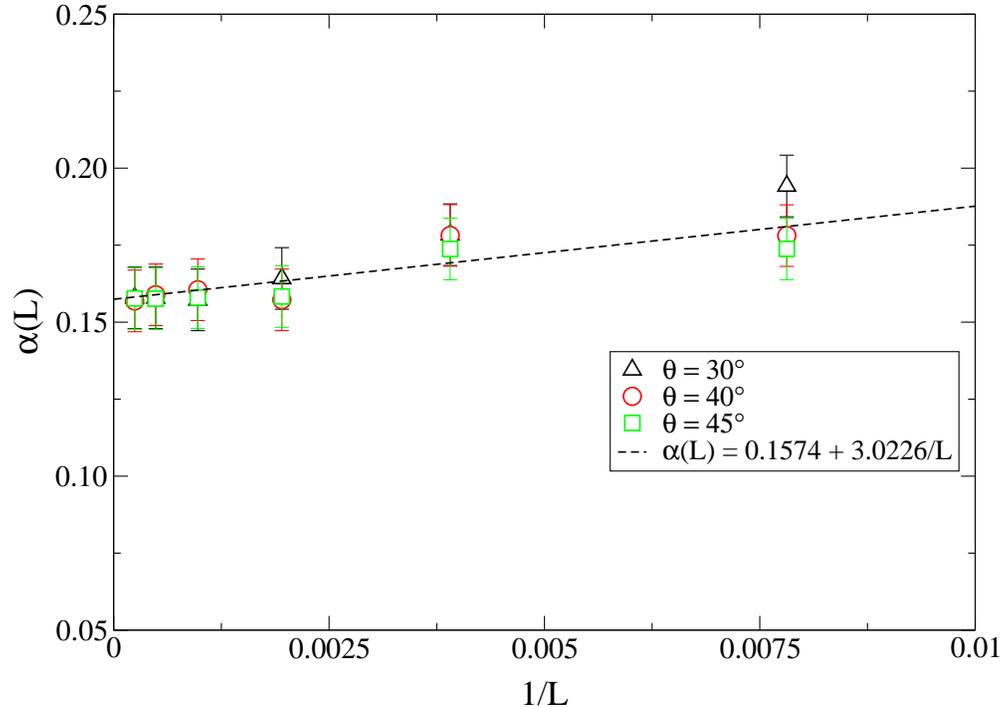}
\caption{Roughness exponent $\alpha$ as function of $L^{-1}$ for different angles of deposition. According to our computer simulations, the best fit we found was $\alpha \approx 0.157$ for large values of L.}
\label{alpha_L}
\end{figure}

We found the best estimative for the roughness exponent was $\alpha \approx 0.157 \pm 0.001$. This estimative is valid for $\theta \geq 20$\textdegree. The plot of $\alpha$ as function of the inverse of the lattice size is presented in the figure~\ref{alpha_L}.

\begin{figure}[!htb]
\includegraphics[width=0.8\linewidth]{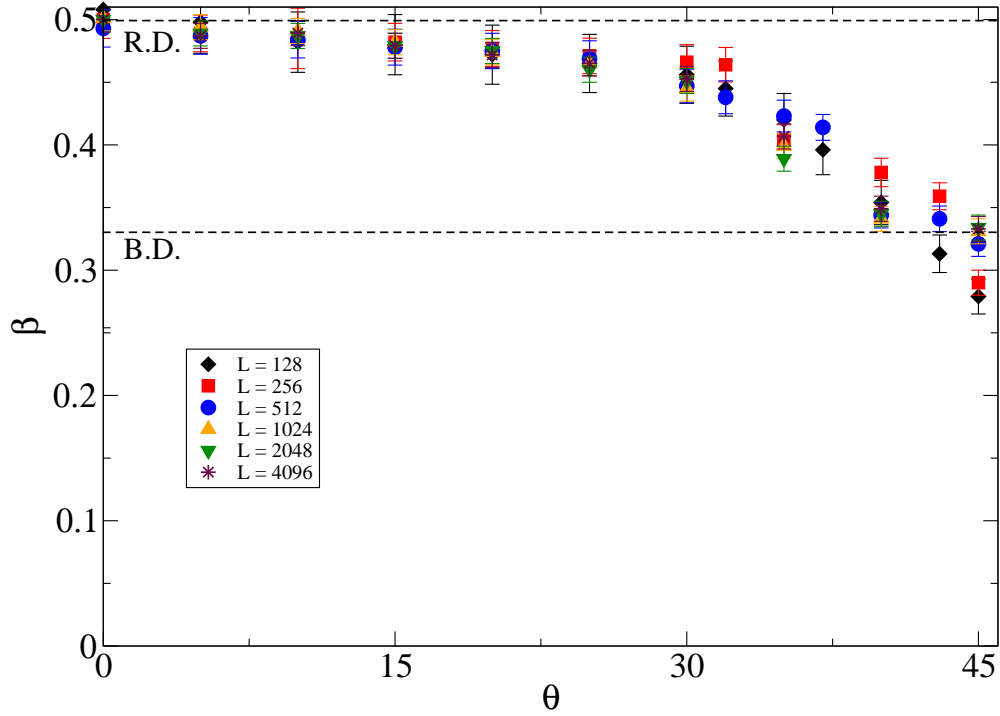}
\caption{Values of the growth exponent $\beta$ as function of the deposition angle $\theta$, for different systems sizes. The dashed lines indicate the values of the growth exponent for the Random Deposition Model and the Ballistic Deposition Model.}
\label{beta_theta}
\end{figure}

Regarding to the growth exponent, as one can see in the figure~\ref{beta_theta}, as the value of $\theta$ increases, the exponent $\beta$ decreases from 1/2 when $\theta = 0$ and becomes closer to the expected value of 1/3 for the Ballistic Deposition Model, when $\theta = 45$\textdegree. For $\theta \geq 45$\textdegree, the value of $\beta$ increases. For the small lattices we obtain higher fluctuations, which is observed for $\theta = 45$\textdegree~in the figure~\ref{beta_theta} for $L \le 512$, where our results do not have a good agreement with the theoretical results. However, for the largest substrates, the simulations have a very good agreement with the expected value of 1/3.

We also analyze our model by means of a stochastic growth equation, in the same form presented in equation \ref{stochastic_eq}. The evolution of the average height of the surface can be described as

\begin{equation}
 \frac{\partial h(\vec{r}, t)}{\partial t} = F + S(f),
\end{equation}
where F represents the average number of particles added to the surface and $S(f)$ represents the random fluctuation of this process, and in this model, a pink noise that will introduce a correlation among the positions on the substrate. The noise has the form

\begin{equation}
 S(f) = (A + Bi)f^{-\gamma},
\end{equation}
where $A$ and $B$ are arbitrary constants, $i$ is the imaginary unit, $f$ is the noise frequency and the exponent $0 \le \gamma \le 2$ is usually close to 1. By the selection of the proper noise and its constants to the stochastic growth equation of the Random Deposition model, one can see that the exact same exponent of the Ballistic Deposition model can be obtained as we will present.

With the appropriated chosen of the constants and using $\gamma = 2/3$, the function $S(f)$ became

\begin{equation}
S(f) = \frac{(-1)^{5/6}\sqrt{\frac{\pi}{2}(sgn(f) -1)}}{\Gamma(\frac{1}{3}) f^{2/3}},
\end{equation}
using the $sgn$ function and the Gamma function, $\Gamma$. As the stochastic equation is a function of time, we use an inverse Fourier transformation in the noise to transform this function from the domain of frequency to the time domain. Using the inverse Fourier transformation,

\begin{equation}
\mathcal{F}(S) = \Phi(\vec{r}, t) = C \frac{1}{t^{1/3}} + D,
\end{equation}
where $t$ is time and $C$ and $D$ are constants. By integration of $\Phi((\vec{r}, t))$ from 0 to $t$,one can write $\langle h(\vec{r}, t) \rangle$, $\langle h^2 \rangle$ and $\langle h \rangle^2$ as

\begin{equation}
\langle h(\vec{r}, t) \rangle = Dt + \int_0^t \Phi(\vec{r}, t) dt = Dt^{2/3}
\end{equation}
which leads to
\begin{equation}
\langle h \rangle^2 = D^2t^{4/3}.
\end{equation}

Using the definition of the surface roughness,
\begin{eqnarray}
w^2(t) = \langle h^2 \rangle - \langle h \rangle^2 \\
w^2(t) = Ct^{2/3} \Rightarrow w(t) = Ct^{1/3},
\end{eqnarray}
and the growth exponent $\beta = 1/3$, the expected value for the Ballistic Deposition. The roughness exponent $\alpha$ cannot be obtained analytically by using the equation \ref{stochastic_eq}, since the simple equation used in this work does not depend on the height $h$. Only through Monte Carlo simulations, for our model, we can obtain the exponent $\alpha$.

\section{Conclusions and further remarks}
\label{Conclusions}

We studied the surface growth due to the deposition of particles dropped at random with different angles of deposition over a liner square lattice using computer simulations and stochastic differential equations. Our model is a modification of the simple RD model in an attempt to make it closer to real deposition processes.

We showed that the surface roughness evolves in time with different behavior, even when $\theta \neq 0$. At initial times, the roughness behaves as in the RD model and $\beta = 0.5$. At long deposition times, the surface roughness grows slowly and the exponent $\beta \approx 0.33$, as observed in the BD model. From our Monte Carlo simulations, we changed the angle of deposition and we showed that when $\theta = 45$\textdegree and the size of the system is large enough, the growth exponent $\beta = 1/3$ is the same of the Ballistic Deposition model.

From our calculations, one can see that the same result from the simulations was obtained by a stochastic equation, using a proper noise - the pink noise - to describe the interface growth and its evolution.

\section*{Acknowledgments}

The authors would like to thank the Federal University of Southern Bahia, UFSB, for the financial support given by the program 005/2015-PROSIS.

\nocite{*}
\bibliographystyle{unsrt}
\bibliography{aipbib}

\end{document}